\documentclass[prb,twocolumn,preprintnumbers,amsmath,amssymb,superscriptaddress]{revtex4}
\usepackage{graphicx}
\usepackage{dcolumn}
\usepackage{bm}
\usepackage{color}
\usepackage{hyperref}
\usepackage{ulem}

\font\scripti=cmmi7
\font\scriptscripti=cmmi5
\def\sib#1{\setbox0 = \hbox{\scripti #1}
  \kern-.02em\copy0\kern-\wd0
  \kern.04em\box0} 
\def\ssib#1{\setbox0 = \hbox{\scriptscripti #1}
  \kern-.02em\copy0\kern-\wd0
  \kern.04em\box0} 
\font\tenib=cmmib10 
\skewchar\tenib='177 \skewchar\tenib='177 \skewchar\tenib='177
\def\pbold#1{\setbox0 = \hbox{$ #1 $}
  \kern-.022em\copy0\kern-\wd0
  \kern.011em\copy0\kern-\wd0
  \kern.011em\copy0\kern-\wd0
  \kern.011em\copy0\kern-\wd0
  \kern.011em\box0} 

\usepackage{graphicx}
\usepackage{dcolumn}
\usepackage{bm}
\usepackage{mathrsfs}

\def\up{\uparrow}
\def\dwn{\downarrow}

\def\lesssim{\ \raise.3ex\hbox{$<$}\kern-0.8em\lower.7ex\hbox{$\sim$}\ }
\def\gesim{\ \raise.3ex\hbox{$>$}\kern-0.8em\lower.7ex\hbox{$\sim$}\ }

\begin{document}
\title{Optical spin transport theory of spin-1/2 topological Fermi superfluids }
\author{Hiroyuki Tajima}
\affiliation{Department of Physics, Graduate School of Science, The University of Tokyo, Tokyo 113-0033, Japan}
\author{Yuta Sekino}
\affiliation{Quantum Hadron Physics Laboratory, RIKEN Nishina Center (RNC), Wako, Saitama, 351-0198, Japan}
\affiliation{Interdisciplinary Theoretical and Mathematical Sciences Program (iTHEMS), RIKEN, Wako, Saitama 351-0198, Japan}
\author{Shun Uchino}
\affiliation{Advanced Science Research Center, Japan Atomic Energy Agency, Tokai, Ibaraki 319-1195, Japan}
\affiliation{Waseda Institute for Advanced Study, Waseda University, Shinjuku, Tokyo 169-8050, Japan}

\date{\today}
\begin{abstract}
We theoretically investigate optical (frequency-dependent) bulk spin transport properties in a spin-1/2 topological Fermi superfluid.
We specifically consider a one-dimensional system with an interspin {\it p}-wave interaction, which can be realized in ultracold atom experiments.
Developing the BCS-Leggett theory to describe the Bardeen-Cooper-Schrieffer (BCS) to Bose-Einstein condensate (BEC) evolution and the $\mathbb{Z}_2$ topological phase transition in this system, we show how the spin transport reflects these many-body aspects. 
We find that the optical spin conductivity, which is a small AC response of a spin current, shows the spin gapped spectrum in the wide parameter region and the gap closes at $\mathbb{Z}_2$ topological phase transition point.
Moreover, the validity of the low-energy effective model of the Majorana zero mode is discussed along the BCS-BEC evolution in connection with the scale invariance at {\it p}-wave unitarity.

\end{abstract}
\maketitle

\section{Introduction}
Topological superconductors and superfluids have accepted special interests from broad communities of modern physics~\cite{qi}.
Phenomena emerging in topological matter such as liquid helium~\cite{mizushima}, unconventional superconductors~\cite{Mackenzie}, $^3P_2$ neutron {superfluids}~\cite{Dean}, and color {superconductors}~\cite{Alford} manifest fascinating properties
relevant to cutting-edge quantum technology.
For instance, Majorana fermions anticipated to appear at the edge of topological systems are crucial key ingredients for topological quantum computation~\cite{Beenakker}.
\par
A spin transport plays a crucial role in revealing topological properties in condensed matter systems such as topological insulators~\cite{qi} and spin Hall systems~\cite{sinova},
where the gapless edge state involves the helical spin transport with the time-reversal symmetry~\cite{KaneMele}.
In particular, properties of AC spin transport provide us with intriguing opportunities to reveal non-trivial aspects of these systems and their application to spintronics~\cite{maekawa}. 
In condensed matter, however, spin transport in mesoscale or submicron scale is  manipulated~\cite{heinrich,woltersdorf,matsuo,jiao,sun,hahn,wei,weiler,li,kobayashi,kurimune}, 
and exploration of bulk spin transport 
in a topological state of matter remains challenging. 
In Ref.~\cite{Sekino2021}, we have recently pointed out that the optical (frequency-dependent) spin conductivity can be measured
in ultracold atomic gases being ideal quantum simulators of condensed matter systems~\cite{schafer,amico,enss-thywissen}.
Since the optical (charge) conductivity spectra have already been measured in an optical lattice system by using the similar method~\cite{Anderson:2019},
detailed examinations of AC spin transport with cold-atom experiments get within reach.
\par
Regarding the realization of topological superfluids in ultracold atomic gases, 
a {\it p}-wave superfluid Fermi gas has been one of the promising candidates for the past few decades~\cite{ohashi:2005,tlho,Gurarie}.
However,
various effects such as three-body loss and dipolar relaxation~\cite{regal,zhang,inada,waseem,Yoshida2018,Waseem2018,Waseem2019} prevent the systems from reaching the superfluid state.
At the same time, it has recently been suggested that such atom loss processes may be suppressed in the low-dimensional systems~\cite{kurlov,zhou,pan,Fonta}.
Shortly afterwards, the corresponding atomic loss measurements in one-dimensional (1D) systems near the {\it p}-wave Feshbach resonance have been performed in several experimental groups~\cite{ytchang,marcum}.
In this regard, a 1D spin-1/2 Fermi gas with an interspin {\it p}-wave interaction{~\cite{Ma:2019,Tajima:pwave}} is one of the possible targets for realizing a topological Fermi superfluid {because its} three-body losses are weak compared to {the fully spin polarized} case, {where the Bose-Fermi duality weakens the Pauli blocking effect in coordinate space at low energy scale~\cite{Cheon:1999,Girardeau:2004,Bender:2005,Girardeau:2006,Girardeau:2008,Cui,Sekino:2018a,Valiente1,Valiente2,sekino:pwave}.}
More explicitly, while a strong {\it p}-wave attraction induces the three-body collision by overwhelming the Pauli-blocking effect in fully polarized case, the Pauli-blocking between two identical fermions can suppress the three-body collision in the present spin-balanced mixture with only interspin interaction.
In addition, the optical spin transport in interacting spin-1/2 systems can be non-trivial even in the absence of a lattice and impurities~\cite{Sekino2021}. 
It is in contrast to {fully spin polarized} fermions, {whose optical spin conductivity becomes independent of interatomic interactions} due to the generalized Kohn's theorem~\cite{wu}.
Moreover, the 1D $p$-wave Fermi gas at {unitarity, where a {\it p}-wave scattering length diverges,} shows the so-called universal thermodynamics~\cite{HoUT} {which makes thermodynamic quantities independent of any length scale} associated with the interaction in spite of the presence of strong correlations~\cite{Bender:2005,Girardeau:2006,Cui,Valiente1,Valiente2,Sekino:2018a,sekino:pwave,Tajima:pwave}. 
Thus, one can investigate a unique interplay between topological and universal aspects of this system, which has never been addressed yet. 
\par
Being motivated by these backgrounds,
we discuss spin transport properties of a 1D unpolarized spin-1/2 {\it p}-wave superfluid Fermi gas at zero temperature.
To this end, we develop the BCS-Leggett theory that allows to describe the BCS-BEC evolution and the topological phase transition in this system.
Using the linear response theory, moreover, we clarify how the optical spin transport properties reflect these non-trivial many-body effects by changing the {\it p}-wave interaction strength.  

In Sec.~\ref{sec2}, we present the formalism of the BCS-Leggett theory and explain topological properties of this system.
In Sec.~\ref{sec3}, we discuss the analytical properties of optical spin transport.
In Sec.~\ref{sec4}, we show the numerical results of bulk thermodynamics and the optical spin conductivity, and discuss the low-energy effective model for the Majorana zero mode.
In Sec.~\ref{sec5}, we summarize this paper.
In what follows, we take $\hbar=k_{\rm B}=1$ and the system size $L$ is taken to be unity.

\section{Theoretical model}
\label{sec2}
\subsection{\label{sec:Hamiltonian}Hamiltonian}
The Hamiltonian for a 1D unpolarized spin-1/2 Fermi gas with interspin {\it p}-wave interaction is given by
\begin{align}
\label{eq:H}
    H&=H_0+V,
\end{align}
where
\begin{align}
\label{eq:H0}
        H_0=\sum_{k,\sigma}\xi_{k}c_{k,\sigma}^\dag c_{k,\sigma}
\end{align}
and
\begin{align}
\label{eq:V}
    V=U\sum_{k,k',q}\Gamma_k\Gamma_{k'}c_{k+q/2,\uparrow}^\dag c_{-k+q/2,\downarrow}^\dag c_{-k'+q/2,\downarrow} c_{k'+q/2,\uparrow}
\end{align}
are {the kinetic term} and the {\it p}-wave interaction term that is assumed to be separable, respectively.
In Eq.~(\ref{eq:H0}), $\xi_k=k^2/(2m)-\mu$ is the kinetic energy with a momentum $k$ measured from the chemical potential $\mu$. 
$c_{k,\sigma}$ is an annihilation operator of a Fermi atom with {spin} $\sigma=\uparrow,\downarrow$.
The coupling constant $U$ is related to the {\it p}-wave scattering length $a$ as
\begin{align}\label{eq:a}
    \frac{m}{2a}=\frac{1}{U}+\sum_{k}\frac{\Gamma_k^2}{2\epsilon_k},
\end{align}
where the form factor $\Gamma_k$ is an odd function of $k$ and $\epsilon_k=k^2/(2m)$.
{In this paper, {we assume $\Gamma_k=O(k)$ for $|k|\to\infty$}, which is justified near a {\it p}-wave resonance in 1D.
Indeed, the form factor in the zero-range limit is given by $\Gamma_k=k$ for any $k$
~\cite{Cui,Valiente1,Valiente2,sekino:pwave,Tajima:pwave,Pastukhov:2020}, while effects of a positive effective range can be taken into account by {the form factors with different shapes 
such as $\Gamma_k=k/(k^2\gamma^2+1)$~\cite{Valiente1}.}
}
We note that while the two-channel model is employed to describe {\it p}-wave Feshbach resonance with a negative effective range in higher dimensions,
one can use the present single-channel model without conflicting with Wigner's causality bound in 1D~\cite{Wigner,Hammer1,Hammer2}.
Also, the parameters of a transverse trapping are included in $a$ in the case of quasi-1D systems{~\cite{Granger:2004,Pricoupenko}.}


\subsection{BCS-Leggett theory}
In a strictly 1D system, superfluid states accompanied by condensation are prohibited by the Mermin-Wagner-Hohenberg theorem~\cite{MW,Hohenberg}.
Nevertheless, here we set the mean-field superfluid state, provided that
the quasi-1D system is concerned where weak three-dimensional properties allow us to describe the quasi-long-range-ordered state within the mean-field approach.
Namely, we introduce the {\it p}-wave superfluid order parameter
\begin{align}\label{eq:Delta}
    \Delta(k)=-U\Gamma_k\sum_{k'}\Gamma_{k'}\langle c_{-k'\downarrow}c_{k'\uparrow}\rangle\equiv \Gamma_kD.
\end{align}
By taking an appropriate gauge transformation, we can take $D$ as a positive value without loss of generality, so that $\Delta(k)$ becomes real valued.
The mean-field Hamiltonian reads
\begin{align}\label{eq:H_MF}
    H_{\rm MF}&=\sum_{k}\Psi_k^\dag
    \left(
    \begin{array}{cc}
        \xi_k &  -\Delta(k)\\
        -\Delta(k) & -\xi_k
    \end{array}
    \right)
    \Psi_{k}-\frac{D^2}{U}+\sum_k\xi_k\cr
    &\equiv \sum_{k}\Psi_k^\dag {H_{\rm BdG}}(k)\Psi_k -\frac{D^2}{U}+\sum_k\xi_{k},
\end{align}
where $\Psi_{k}=(c_{k,\uparrow} \ c_{-k,\downarrow}^\dag)^{\rm T}$ is the two-component Nambu spinor.
Although the mean-field Hamiltonian for the spin-triplet superfluid is generally described in terms of the four-component Nambu spinors,
we do not have to use the four-component one since the present system involves only the interspin $p$-wave pairing interaction and the off-diagonal part for the equal spin pairing ($\uparrow\uparrow$ and $\downarrow\downarrow$) is trivially zero.
The Bogoliubov transformation
\begin{align}
\label{eq:bog}
    \left(\begin{array}{c}
         \alpha_{k,1}  \\
          \alpha_{k,2}^{{\dag}}
    \end{array}\right)
    &=\left(
    \begin{array}{c}
         u_k c_{k,\uparrow}-v_k c_{-k,\downarrow}^\dag  \\
         u_k c_{-k,\downarrow}^\dag +v_{k} c_{k,\uparrow}
    \end{array}
    \right)
\end{align}
leads to
\begin{align}
    H_{\rm MF}&={\sum_{k}\sum_{i=1,2}E_k\alpha_{k,i}^\dag \alpha_{k,i}+E_{\rm GS},}
\end{align}
where
\begin{align}
    E_k=\sqrt{\xi_k^2+\Delta^2(k)}{=}\sqrt{\xi_k^2+D^2\Gamma_k^2}
\end{align}
is the dispersion of the Bogoliubov quasiparticle and
\begin{align}
    E_{\rm GS}=-\frac{D^2}{U}+\sum_k\left(\xi_{k}-E_{k} \right)
\end{align}
is the ground-state energy.
{In Eq.~(\ref{eq:bog}), $u_k^2=\frac{1}{2}(1+\xi_k/E_k)$
and $v_k^2=\frac{1}{2}(1-\xi_k/E_k)$ are the BCS coherence factors.}
{For given $a$ and the particle number $N$, $D$ and $\mu$ are determined by solving the following two equations self-consistently: The first one is the so-called gap equation,
\begin{align}
\label{eq:gapeq}
    \frac{m}{2a}+\sum_{k}\Gamma_k^2\left[\frac{1}{2E_k}-\frac{1}{2\varepsilon_k}\right]=0,
\end{align}
resulting from the minimization condition of $E_{\rm GS}$ with respect to $D$, while the other one is the particle number equation
\begin{align}
\label{eq:Neq}
    N=-\frac{\partial E_{\rm GS}}{\partial \mu}=\sum_{k}\left[1-\frac{\xi_k}{E_k}\right].
\end{align}
Discussions of numerically evaluated $D$ and $\mu$ are presented in Sec.~\ref{sec:numerics} (see Fig.~\ref{fig:1}).}
{We note that a mean-field theory for spin polarized Fermi atoms with {\it p}-wave interaction has been studied in a similar way~\cite{Pastukhov:2020}.}

\subsection{Topological classification}
Here we revisit the classification of topological superconductors{/superfluids} and show the symmetry class of the present system~\cite{Sato}.
The Bogoliubov-de Genne (BdG) Hamiltonian $H_{\rm BdG}(k)$ in Eq.~(\ref{eq:H_MF}) can be rewritten as
\begin{align}
\label{eq:Hbdg}
    H_{\rm BdG}(k)=\bm{\sigma}\cdot \bm{R}(k),
\end{align}
where $\bm{\sigma}=(\sigma_x,\sigma_y,\sigma_z)$ is {a set of the Pauli matrices} acting on the particle-hole space, and we have defined
\begin{align}
\bm{R}(k)=(-\Delta(k),0,\xi_k)
\end{align}
We note that the absence of $\sigma_y$ terms in Eq.~(\ref{eq:Hbdg}) results from the real-valued $\Delta(k)$.
For classification of topological superfluids/superconductors, we start to check the particle-hole-like symmetry for $H_{\rm BdG}(k)$ given by the following relation:
\begin{align}\label{eq:PHS}
    \Xi^{-1}H_{\rm BdG}(-k)\Xi=-H_{\rm BdG}(k)
\end{align}
where $\Xi=\sigma_xK$ is the charge-conjugation-like operator with $\Xi^2=1$ and $K$ is a complex-conjugate operator.
In addition, since $H_{\rm BdG}(k)$ anticommutes with $\sigma_y$, $H_{\rm BdG}(k)$ has the chiral symmetry given by
\begin{align}\label{eq:CS}
C^{-1}H_{\rm BdG}(k)C=H_{\rm BdG}(k),
\end{align}
where $C=i\sigma_y$ is the chiral operator.
The chiral operator is related to $\Xi$ and the time-reversal-like operator $\Theta$ as $C=\Theta\Xi$.
We can find $\Theta=\sigma_z K$ and the following time-reversal-like symmetry:
\begin{align}\label{eq:TRS}
\Theta^{-1}H_{\rm BdG}(-k)\Theta=H_{\rm BdG}(k).
\end{align}
Equations~(\ref{eq:PHS}), (\ref{eq:CS}), and (\ref{eq:TRS}) combined with $\Xi^2=\Theta^2=+1$ show that our 1D superfluid belongs to the class BDI~\cite{Schnyder}.
The $\mathbb{Z}$ topological invariant characterizing this system is given by the winding number
\begin{align}\label{eq:winding_number}
\nu=\int_{-\infty}^\infty\frac{dk}{2\pi i}q^*(k)\frac{\partial}{\partial k}q(k),
\end{align}
where $q(k)=\hat{R}_z(k)-i\hat{R}_x(k)$ and $\hat{\bm{R}}(k)=\bm{R}(k)/|\bm{R}(k)|$.
We note that the $\mathbb{Z}_2$ topological invariant ${\nu_2}$ defined by
$(-1)^{\nu_2}={\rm sgn}[\hat{R}_z(k=0)]{\rm sgn}[\hat{R}_z(k\rightarrow\infty)]$ corresponds to the parity of $\nu$.
{As mentioned in Sec.~\ref{sec:Hamiltonian}, $\Gamma_k=\Delta(k)/D$ satisfies 
$\Gamma_k=O(k)$ for $k\rightarrow\infty$, leading to
\begin{align}\label{eq:nu}
(-1)^{\nu_2}={\rm sgn}(-\mu).
\end{align}} 
While the case of $\nu_2=1$, $\mu>0$ corresponds to the mapping to the trajectory from the south pole $\hat{\bm{R}}(k=0)=(0,0,-1)$ to the north pole $\hat{\bm{R}}(k\rightarrow \infty)=(0,0,1)$ with increasing $k\geq 0$,
the other case ($\nu_2=0$, $\mu<0$) does to the trivial trajectory{, where both the starting and ending points are the north pole.}

{We emphasize that the above-mentioned classification of phases and the topological invariant based on the sign of $\mu$ [Eq.~(\ref{eq:nu})] is valid regardless of details of a resonance such as an effective range.
Since the key of the above discussion is $\Delta(k)/\xi_k\to0$ for $|k|\to\infty$ and $|k|\to0$ resulting from the assumptions of $\Gamma_k$, the discussion holds both in the zero-range limit and in the presence of a positive effective range.
In addition, Eq.~(\ref{eq:nu}) is also valid in the case with a negative effective range, where the mean-field order parameter $\Delta(k)$ is proportional to $k$~\cite{Ohashi:2005,Gurarie:2005,Gurarie:2006}.
Hereafter, we take the zero-range case $\Gamma_k=k$ for simplicity.
In this case, we obtain $\nu_2=\nu$.
The systems with $\mu>0$ ($\mu<0$) has the topological invariant $\nu=1$ ($\nu=0$), and
the $\mathbb{Z}_2$ topological phase transition occurs at $\mu=0$.
}


\section{Optical spin transport}
\label{sec3}
In this section, we analytically evaluate the optical spin conductivity in a spin-1/2 {\it p}-wave topological superfluid at $T=0$.
On the basis of our previous paper~\cite{Sekino2021}, we consider 1D fermions under  a small external spin-dependent force $F_{S}(t)$.
The corresponding perturbative Hamiltonian is given by $\delta H(t)=-\int dx F_{S}(t)xS(x)$, where $S(x)=[\psi_\up^\dagger(x)\psi_\up(x)-\psi_\dwn^\dagger(x)\psi_\dwn(x)]/2$ with $\psi_\sigma(x)=\sum_k c_{k,\sigma}e^{ikx}$ is the local spin imbalance.
By monitoring the spin-selective center-of-mass motion under the external spin-dependent force $F_S(\omega)$ with the frequency $\omega$ (which is the Fourier transform of $F_S(t)$),
one can measure the optical spin conductivity $\sigma^{(S)}(\omega)=\langle J_S(\omega)\rangle/\tilde{F}_S(\omega)$ where $\langle J_S(\omega)\rangle$ is the Fourier transform of the thermal average of the spin current operator $J_S(t)= \frac{d}{dt}\int dx S(x,t)x$
(see Ref.~\cite{Sekino2021} for more details).
{The linear response theory relates $\sigma^{(S)}(\omega)$ to the retarded response function $\chi(\omega)$ of a spin current as}
\begin{align}\label{eq:sigma}
\sigma^{(S)}(\omega)=\frac{i}{\omega_+}
\left[\frac{N}{4m}+\chi(\omega)\right],
\end{align}
where $\omega_+=\omega+i\eta$ with an infinitesimal positive number $\eta$.
%
%
$\chi(\omega)$ is defined as
\begin{align}
    \chi(\omega)=-i\int_0^{\infty}dt e^{i\omega_+t}
    \langle [J_{S}(t),J_S(0)]\rangle
\end{align}
and $\langle\cdots\rangle$ denotes the expectation value with respect to the ground state.
The response function in the BCS-Leggett theory can be evaluated in the same way as in the case of a 3D {\it s}-wave superfluid Fermi gas~\cite{Sekino2021}:
\begin{align}
\label{eq:chi}
    \chi(\omega)=\sum_{k}\frac{k^4D^2}{4m^2E_k^2}
    \left(\frac{1}{\omega_+-2E_k}-\frac{1}{\omega_++2E_k}\right).
\end{align}
{
Using Eqs.~(\ref{eq:Neq}) and (\ref{eq:chi}), one can confirm that the optical spin conductivity satisfies the $f$-sum rule~\cite{Sekino2021,Enss2}}
\begin{align}
    \int_{-\infty}^{\infty}d\omega\,{\rm Re}\left[\sigma^{(S)}(\omega)\right]
    &=\frac{\pi}{4m}N.
\end{align}

{Since the imaginary part of the optical spin conductivity can be expressed in terms of the real part with the Kramers-Kronig relation, we hereafter focus only on ${\rm Re}\left[\sigma^{(S)}(\omega)\right]$.
From Eq.~(\ref{eq:sigma}), we obtain
\begin{align}
\label{eq:chid}
    {\rm Re}\left[\sigma^{(S)}(\omega)\right]
    &=\mathcal{D}_S\delta(\omega)-\frac{1}{\omega}{\rm Im}\chi(\omega),
\end{align}
where the spin Drude weight
\begin{align}
\mathcal{D}_S&=\pi\left[\frac{N}{4m}+{\rm Re}\chi(0)\right]
\end{align}
characterizes the sharp contribution at zero frequency.
Using Eq.~(\ref{eq:chi}) as well as Eq.~(\ref{eq:Neq}), one can find that the spin Drude weight in this superfluid always vanishes ($\mathcal{D}_S=0$).
Substituting Eq.~(\ref{eq:chi}) into the second term in Eq.~(\ref{eq:chid}) yields
\begin{align}\label{eq:Re_sigma}
    {\rm Re}\left[\sigma^{(S)}(\omega)\right]=\sum_{k}\frac{\pi k^2\Delta^2(k)}{m^2|\omega|^3}\delta(|\omega|-2E_k).
\end{align}
This equation indicates that the spectrum of ${\rm Re}\left[\sigma^{(S)}(\omega)\right]$ is sensitive to the shape of the quasiparticle dispersion $E_k=\sqrt{\xi_k^2+D^2k^2}$.
In particular, ${\rm Re}\left[\sigma^{(S)}(\omega)\right]$ vanishes for $|\omega|$ below the spin gap
$E_{\rm gap}={\rm min}[2E_{k}]$.
For $\mu\geq mD^2$, $E_{k}$ becomes minimum at a nonzero momentum, while, for $\mu <mD^2$, $E_{k}$ monotonically increases with increasing $|k|$, leading to
\begin{align}
\label{eq:emin}
    E_{\rm gap}
    =\left\{
    \begin{array}{cc}
     2D\sqrt{2m\mu-m^2D^2}     &\mu> mD^2 \\
     2|\mu|     &\mu <mD^2
     \end{array}
    \right.,
\end{align}
which shows that the spin gap is closed at $\mu=0$ corresponding to the $\mathbb{Z}_2$ topological phase transition point.

The value of the chemical potential characterizes not only the topological phases [see Eq.~(\ref{eq:nu})] but also the gap structures [Eq.~(\ref{eq:emin})] in the quasiparticle dispersion.
For this reason, we define three regions of $\mu$ with nonzero spin gap; (i) $\mu> mD^2$, (ii) $0<\mu<mD^2$, and (iii) $\mu <0$.
The regions (i) and (ii) [(iii)] are in the topologically non-trivial (trivial) phase with $\nu=1$ ($\nu=0$), and in the region (i) [(ii) and (iii)] the superfluid has the spin gap associated with the nonzero-momentum (zero-momentum) quasiparticle excitation.}

Performing the momentum summation in Eq.~(\ref{eq:Re_sigma}), we obtain an analytical expression of ${\rm Re}[\sigma^{(S)}(\omega)]$ as
\begin{align}
\label{eq:sig}
&{\rm Re}\left[\sigma^{(S)}(\omega)\right]
=\frac{D\theta(|\omega|-E_{\rm gap})}{4m|\omega|^2\sqrt{m^2D^2-2m\mu+\left(\frac{|\omega|}{2D}\right)^2}}\cr
&\quad\times \left[\mathcal{K}_{+}^3(|\omega|)+\mathcal{K}_{-}^3(|\omega|)\theta(\mu-mD^2)\theta(2|\mu|-|\omega|)\right],
\end{align}
where we have defined
\begin{align}
    &\mathcal{K}_{\pm}^2(|\omega|)=2m(\mu-mD^2)\cr
    &\quad \pm 2mD\sqrt{m^2D^2-2m\mu+\left(\frac{|\omega|}{2D}\right)^2}.
\end{align}
{In the region (i) with $\mu> mD^2$, the spectrum of ${\rm Re}[\sigma^{(S)}(\omega)]$ shows the coherence peak ${\rm Re}\left[\sigma^{(S)}(\omega)\right]\sim1/\sqrt{\omega-E_{\rm gap}}$ for $\omega\to E_{\rm gap}+0$, while, in the regions (ii) and (iii), the conductivity {monotonically vanishes in such a limit} without exhibiting a coherence peak.
It should be noted that the step function $\theta(|\omega|-E_{\rm gap})$ in the numerator of Eq.~(\ref{eq:sig}) indicates that
${\rm Re}\left[\sigma^{(S)}(\omega)\right]$ vanishes below the spin gap $E_{\rm gap}$ given by Eq.~(\ref{eq:emin}).
We will discuss this difference in spin conductivity spectra for various interaction strengths in the next section (see Fig.~\ref{fig:2}).
In the high-frequency limit, ${\rm Re}[\sigma^{(S)}(\omega)]$ has the following power-law tail:}
\begin{align}
\label{eq:hft}
    \lim_{\omega\rightarrow \infty}{\rm Re}\left[\sigma^{(S)}(\omega)\right]&=\frac{C}{4(m|\omega|)^{\frac{3}{2}}},
\end{align}
where $C=2m^2D^2$ is the {\it p}-wave contact obtained from
the adiabatic theorem~\cite{Cui}
\begin{align}
    \frac{\partial E_{\rm GS}}{\partial a^{-1}}=-\frac{C}{4m}.
\end{align}

We examine the optical spin conductivity at the $\mathbb{Z}_2$ topological phase transition point with $\mu=0$.
{In this case, the gapless excitation of the quasiparticle makes the spin gap in ${\rm Re}\left[\sigma^{(S)}(\omega)\right]$ closed.
Equation~(\ref{eq:sig}) reduces to
\begin{align}
{\rm Re}\left[\sigma^{(S)}(\omega)\right]
&=\frac{mD^3\left[\sqrt{1+\omega^2/(4m^2D^4)}-1\right]^{3/2}}{\sqrt{2}\omega^2\sqrt{1+\omega^2/(4m^2D^4)}}
\end{align}
for any $\omega$.
In particular, the spin conductivity linearly behaves in a small frequency region:}
\begin{align}
\label{eq:glc}
    {\rm Re}\left[\sigma^{(S)}(\omega)\right]
    =\frac{|\omega|}{32m^2D^3}+O(|\omega|^2).
\end{align}
We emphasize that {this gapless behavior in the spectrum of the optical spin conductivity is clearly different from that of the conventional Drude-type conductivity with a sharp peak at low frequency.}

\section{Results and Discussions}
\label{sec4}
\subsection{\label{sec:numerics}Numerical results}
\begin{figure}[t]
    \centering
    \includegraphics[width=7.cm]{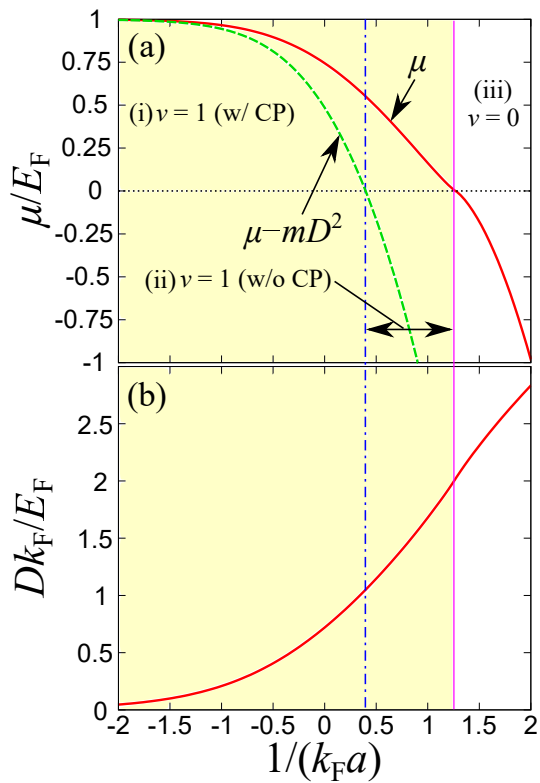}
    \caption{Calculated (a) chemical potential $\mu$ and (b) gap parameter $D$ along the {\it p}-wave BCS-BEC evolution with increasing {the interaction strength $1/(k_{\rm F}a)$.}
    $E_{\rm F}$ and $k_{\rm F}$ are the Fermi energy and momentum, respectively.
    The $\mathbb{Z}_2$ topological phase transition occurs at $1/(k_{\rm F}a)=1.27$ where $\mu=0$. On the other hand, the coherence peak (CP) in the optical spin conductivity disappears at $\mu-mD^2=0${, which is different from the 3D {\it s}-wave case where CP disapears at $\mu=0$}.}
    \label{fig:1}
\end{figure}
Figure~\ref{fig:1}
shows the chemical potential $\mu$ and the gap parameter $D$ along the {\it p}-wave BCS-BEC evolution obtained by solving Eqs.~(\ref{eq:gapeq}) and (\ref{eq:Neq}),
where $k_{\rm F}=\frac{\pi N}{2}$ and $E_{\rm F}=\frac{k_{\rm F}^2}{2m}$ are the Fermi momentum and Fermi energy of a non-interacting Fermi gas, respectively.
While $\mu$ is equal to $E_{\rm F}$ in the weak-coupling limit {[$1/(k_{\rm F}a)\to-\infty$]}, $\mu$ decreases with increasing the interaction strength {$1/(k_{\rm F}a)$} and finally changes its sign at $1/(k_{\rm F}a)={4/\pi=}1.27$, where the $\mathbb{Z}_2$ topological phase transition occurs from {the non-trivial to trivial phases} ($\nu=1\rightarrow 0$).
Simultaneously, $D$ monotonically increases with increasing $1/(k_{\rm F}a)$.
In contrast to $\mu$, $D$ does not exhibit a kink at  $1/(k_{\rm F}a)=1.27$ because the absolute value of $D$ is not important for the topological transition unless $D=0$.
These interaction dependencies of $\mu$ and $D$ are similar to those of the {\it s}-wave BCS-BEC crossover in 3D~\cite{Zwerger,Randeria,Strinati,Ohashi} 
{(although the topological phase transition is absent in the latter case).
{We note that at $\mu=0$, one can analytically solve Eqs.~(\ref{eq:gapeq}) and (\ref{eq:Neq}) as $D=v_{\rm F}$ and $1/(k_{\rm F}a)=4/\pi$, where $v_{\rm F}=k_{\rm F}/m$ is the Fermi velocity.}
On the other hand, the region of $\mu$ with the coherence peak is different between our {\it p}-wave case and the 3D {\it s}-wave case.
As shown in the previous section, $\mu>mD^2$ [the region (i)] corresponds to the case with the coherence peak for the {\it p}-wave superfluid, while $\mu>0$ does in the 3D system~\cite{Zwerger,Randeria,Strinati,Ohashi} .
In Fig.~\ref{fig:1}, we also plot $\mu-mD^2$ as a function of the interaction strength and one can see that $\mu=mD^2>0$ is satisfied at $1/(k_{\rm F}a)=(4/\pi)\left[1+4 \Gamma \left(\frac{5}{4}\right)^2/\Gamma \left(\frac{3}{4}\right)^2\right]^{-1}=0.399$, where $\Gamma(z)$ is the gamma function.
}

\begin{figure}[t]
    \centering
    \includegraphics[width=7.5cm]{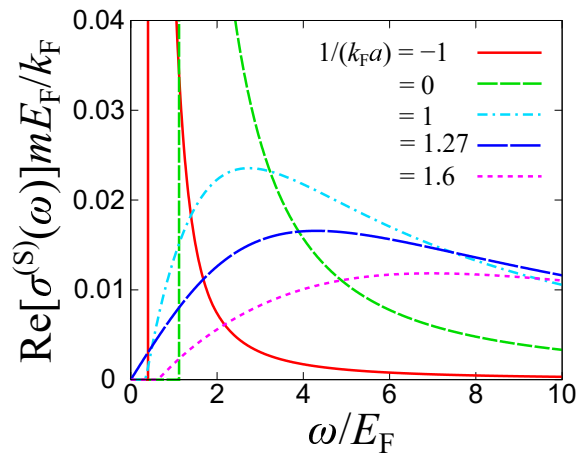}
    \caption{Optical spin conductivity ${\rm Re}\left[\sigma^{(S)}(\omega)\right]$ at various interaction strengths.}
    \label{fig:2}
\end{figure}
Using Eq.~(\ref{eq:sig}) combined with the results of $\mu$ and $D$ shown in Fig.~\ref{fig:1},
we plot the real part of the optical spin conductivity ${\rm Re}[\sigma^{(S)}(\omega)]$ at various interaction strength in Fig.~\ref{fig:2}.
{In the cases of a weak coupling [$1/(k_{\rm F}a)=-1$] and the {\it p}-wave unitarity [$1/(k_{\rm F}a)=0$], the system belongs to the region (i) with $\nu=1$, and ${\rm Re}[\sigma^{(S)}(\omega)]$ exhibits the spin gap and the coherence peak.
On the other hand, the coherence peak in ${\rm Re}[\sigma^{(S)}(\omega)]$ disappears at $1/(k_{\rm F}a)=1$ [the region (ii)].
In this case, the system remains spin gapped and has the same topological invariant ($\nu=1$) as in the weaker coupling side.
One can find the closing of the spin gap at $1/(k_{\rm F}a)=1.27$, where the $\mathbb{Z}_2$ topological phase transition occurs.
Finally, at stronger coupling [$1/(k_{\rm F}a)=1.6$], 
${\rm Re}[\sigma^{(S)}(\omega)]$ shows the spin gap again, indicating that the system undergoes the topologically trivial phase, i.e., the region (iii) with $\nu=0$.
}

\begin{figure}[t]
    \centering
    \includegraphics[width=7.5cm]{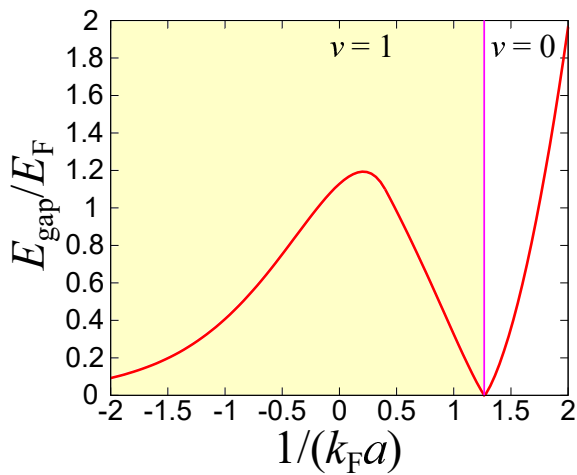}
    \caption{Spin-gap energy $E_{\rm gap}$ {[Eq.~(\ref{eq:emin})]} in the optical spin conductivity ${\rm Re}[\sigma^{(S)}(\omega)]$, as a function of the interaction strength $1/(k_{\rm F}a)$.
    The vertical line at $1/(k_{\rm F}a)=1.27$ indicates the  gap-closing point accompanying the topological phase transition.}
    \label{fig:3}
\end{figure}
Figure~\ref{fig:3} shows the spin-gap energy $E_{\rm gap}$ in ${\rm Re}[\sigma^{(S)}(\omega)]$.
Indeed, one can find $E_{\rm gap}=0$ at $1/(k_{\rm F}a)=1.27$, where the $\mathbb{Z}_2$ topological phase transition occurs, and $E_{\rm gap}>0$ {away from the transition point.}
By recalling the analytical form of the spin-gap energy given by Eq.~(\ref{eq:emin}), $E_{\rm gap}$ is proportional to $|\mu|$ around the transition point.
Moreover, interestingly, $E_{\rm gap}$ exhibits a local maximum around $1/(k_{\rm F}a)=0.399$, 
where the coherence peak disappears (see Fig.~\ref{fig:1}).
Indeed, the analytical form of $E_{\rm gap}$ [Eq.~(\ref{eq:emin})] changes at $\mu=mD^2$.
This result implies that the fermionic character of {\it p}-wave superfluidity qualitatively changes to that of the molecular bosonic condensates in this regime without any phase transitions as usual BCS-BEC crossover phenomena.

\par
\begin{figure}[t]
    \centering
    \includegraphics[width=8cm]{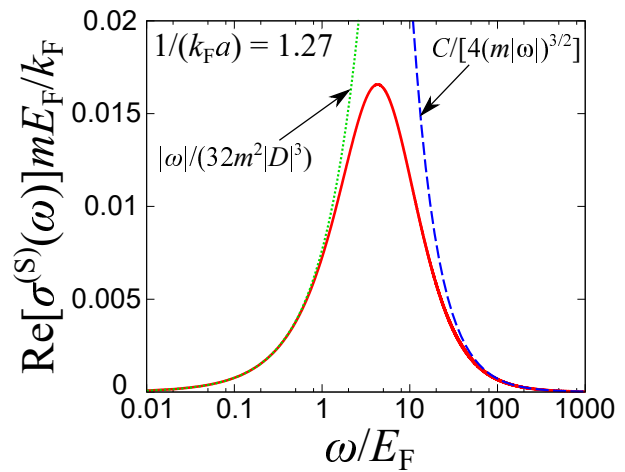}
    \caption{The optical spin conductivity ${\rm Re}[\sigma^{(S)}(\omega)]$ at the gapless point with $\mu=0$ and $1/(k_{\rm F}a)=1.27$.  
    {The dotted and dashed lines denote the asymptotic behaviors [Eqs.(\ref{eq:hft}) and (\ref{eq:glc})] at high and low frequencies, respectively.}}
    \label{fig:4}
\end{figure}

We are now in the position to examine the detailed structure of the optical spin conductivity.
Figure~\ref{fig:4} shows the frequency dependence of ${\rm Re}[\sigma^{(S)}(\omega)]$ at the topological phase transition point $1/(k_{\rm F}a)=1.27$.
One can confirm the linear behavior shown in Eq.~(\ref{eq:glc}) in the sufficiently low-frequency regime.
{This means that the measurement of the optical spin conductivity can detect the $\mathbb{Z}_2$ topological phase transition from the gapless behavior of ${\rm Re}[\sigma^{(S)}(\omega)]$}.
This is analogous to the spin superfluidity at the phase boundary of the spinor Bose condensates~\cite{Sekino2021}, whereas the linear spin conductivity spectra is proportional to the inverse spin velocity $v_{\rm s}^{-1}$ in this bosonic system.
In the present case at $\mu=0$, the dispersion reads
\begin{align}
E_k=\sqrt{\epsilon_k^2+D^2k^2}=D|k|+\frac{|k|^3}{8m^2D}+O(|k|^5),
\end{align}
indicating the gapless spin excitation with {the spin velocity $v_{\rm s}={D\equiv v_{\rm F}}$.}
Since the Drude-like conductivity is generally a decreasing function of $\omega$ in the low-frequency regime,
one can clearly distinguish this gapless spectrum from the Drude one by confirming the linearly increasing behavior of ${\rm Re}[\sigma^{(S)}(\omega)]$.
{In addition, the fact that a vanishing chemical potential results in a gapless linear behavior in ${\rm Re}[\sigma^{(S)}(\omega)]$ remains correct even in the presence of the effective-range corrections.}
Such a low-frequency gapless behavior is in contrast to the s-wave superfluid case where ${\rm Re}[\sigma^{(S)}(\omega)]$ is always gapped~\cite{Sekino2021}.
In the high-frequency regime, the low-energy {spin excitation} becomes irrelevant and ${\rm Re}[\sigma^{(S)}(\omega)]$ exhibits the high-frequency tail being proportional to the {\it p}-wave contact $C$.
The same behavior has also been reported in other systems such as the 3D {\it s}-wave unitary Fermi gas~\cite{Sekino2021,Enss1,Enss2} and spinor BEC~\cite{Sekino2021}.
We note that the high-frequency tail in ${\rm Re}[\sigma^{(S)}(\omega)]$ appears in the entire {\it p}-wave BCS-BEC evolution shown in Fig.~\ref{fig:2}. 
\par
\subsection{Low-energy effective model for the Majorana zero mode at the edge of gas cloud}
\begin{figure}[t]
    \centering
    \includegraphics[width=8cm]{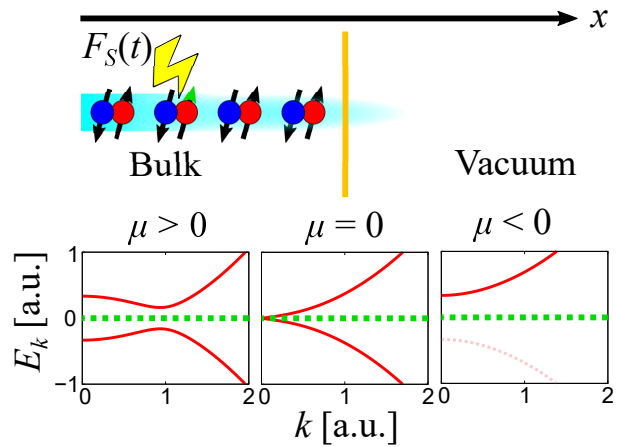}
    \caption{Schematic figure for proving the optical spin conductivity in 1D spin-1/2 topological {\it p}-wave superfluidity.
    The lower panels show the energy dispersion $E_K$ for $\mu>0$, $\mu=0$, and $\mu<0$.
    The Majorana zero mode exits where the local chemical potential becomes zero, $\mu(x)=0$.
    {We note that the negative dispersion $-E_k=-\sqrt{\left(\frac{k^2}{2m}+|\mu|\right)^2+D^2k^2}$ appears in the BEC side with $a^{-1}>0$ and $\mu<0$.
    Physically, this branch describes the hole-like excitation when breaking the tightly-bound $p$-wave molecule.
    Therefore, such a dispersion is absent in vacuum without two-body bound states. 
    }
    }
    \label{fig:edge}
\end{figure}

Here we consider the low-energy effective Hamiltonian for the Majorana zero mode~\cite{Sato2} in the present spin-1/2 {\it p}-wave superfluid system.
The BdG Hamiltonian in the momentum space [{Eq.~(\ref{eq:Hbdg})}] can be rewritten as
\begin{align}
    H_{\rm BdG}(k)
    &=Dk\sigma_x-\mu\sigma_z+O(k^2).
\end{align}
In this regard, the low-energy effective Hamiltonian {density} reads
\begin{align}
\label{eq:heff}
    {\mathcal{H}_{\rm eff}(x)}=-iD(x)\sigma_x\partial_x -\mu(x)\sigma_z,
\end{align}
which is a Dirac Hamiltonian in $(1+1)$ dimensions, and thus
the Majorana zero mode appear at $\mu(x)=0$~\cite{Sato2}.
While we have discussed the bulk optical spin transport,
the Majorana edge state can be detected by measuring the local optical spin conductivity where the local spin-dependent drive is applied as schematically shown in Fig.~\ref{fig:edge}.
\par
Hereafter, we discuss the robustness of the low-energy effective model at {\it p}-wave unitarity even in the absence of the proximity effect, as a result of an interplay between the topological properties and the universal thermodynamics.
To see this, we consider the density dependence of $D$ and $\mu$.
At $a^{-1}=0$, these quantities are scale invariant and exhibit 
\begin{align}
    D\propto \frac{E_{\rm F}}{k_{\rm F}}\sim n^{1}, \quad \mu\propto E_{\rm F} \sim n^2,
\end{align}
where $n=N/L$ is the number density.
Thus, at zero density limit ($n\rightarrow 0$) corresponding to the edge region,
we can safely obtain the gapless Hamiltonian at $\mu(x^*)=0$ as 
\begin{align}
    \mathcal{H}_{\rm eff}(x^*)=-iD(x^*)\sigma_x\partial_x + O(n^2)
    \end{align}
for the Majorana edge mode.
Also, in the BEC side ($a^{-1}>0$), this effective model works well since a nonzero $D$ is obtained even at $\mu=0$ in the bulk system.
However, in such a case, the Majorana edge mode does not appear at the edge of gas cloud but it does around the dilute region of cloud where $\mu(x)=0$ because the local density can be nonzero even for $\mu<0$.

On the other hand, the validity of the effective Hamiltonian {[Eq.~(\ref{eq:heff})]} is not guaranteed in the BCS side $a^{-1}<0$.
The density dependence of $D$ and $\mu$ in the BCS side is given by
\begin{align}
    D\sim n e^{-\frac{1}{|a|n}}, \quad \mu\rightarrow  E_{\rm F}\propto n^2.
\end{align}
Because of non-universal effects associated with finite $a$, $D$ becomes exponentially small in the dilute limit ($n\rightarrow 0$).
In such a case, the decrease of $D$ is faster than that of $\mu$. Moreover, the higher derivative term becomes non-negligible.
Although the magnitude of $D$ around the edge region is assumed to be enough large due to the proximity effect, 
such an induced gap should be small in the weak-coupling regime.
Therefore, we find that the low-energy description of the Majorana zero mode at the cloud edge based on Eq.~(\ref{eq:heff}) is more robust around the {\it p}-wave unitarity limit compared to the BCS regime.
This is a special feature due to the scale invariance at {\it p}-wave unitarity.
\par
{We note that the above discussion provides the fragility of the derivative expansion given by Eq.~(\ref{eq:heff}) in the BCS regime.
Although such an effective theory is broken down in the weak coupling regime, the existence of the Majorana zero mode would be investigated by solving the BdG equation
in a similar way as for 2D trapped chiral {\it p}-wave Fermi superfluids~\cite{Mizushima:2008}.}

\section{Summary}
\label{sec5}
To summarize, we have theoretically investigated optical spin transport properties in spin-1/2 topological {\it p}-wave superfluidity in 1D, which is one of the promissing candidates for realizing topological superfluid Fermi gases in recent cold atom experiments.
\par
We have extend the BCS-Leggett theory for 3D {\it s}-wave BCS-BEC crossover phenomena to the 1D {\it p}-wave BCS-BEC evolution accompanying the $\mathbb{Z}_2$ topological phase transition at zero temperature.
We have introduced the mean-field model Hamiltonian and how to characterize {\it p}-wave interaction with the {\it p}-wave scattering length in this system.
Also, we have clarified that the present 1D continuum system belongs to the symmetry class BDI.
{We have found that topological characterization with chemical potential [Eq.~(\ref{eq:nu})] holds not only in the zero-range limit but also in the presence of effective-range corrections.}
\par
Combining the BCS-Leggett theory and the linear response approach, we have derived the analytical formula of the optical spin conductivity along the {\it p}-wave BCS-BEC evolution.
The optical spin conductivity shows the spin-gapped spectrum at various interaction strengths {away from} the topological phase transition point with the vanishing chemical potential.
On the basis of optical spin transport properties, we have classified three regimes, that is, (i) topologically non-trivial phases with the coherence peak in the BCS side, (ii) topologically non-trivial phase without the coherence peak, and (iii) topologically trivial phase in the BEC side.
Moreover, {the gapless linear behavior in the optical spin conductivity spectrum} at the topological phase transition point is found to be distinct from the {conventional} Drude-type conduction.
{The measurement of the optical spin conductivity, therefore, can detect the topological phase transition as the closing of the spin gap.}
Finally, we have argued the low-energy effective Hamiltonian for the Majorana zero mode.
We have showed that the scale invaliance at {\it p}-wave unitarity assists the low-energy description based on the derivative expansion even in the absence of the proximity effect.
\par
For future work,
it is interesting to investigate how the gapped optical spin conductivity changes to the Drude-type conductivity at finite temperature.
For instance, the spin Drude weight can be nonzero at finite temperature.
The spin-gapped behavior would also remain above the superfluid critical temperature $T_{\rm c}$ due to the emergence of the pseudogap associated with pairing fluctuations.
Such a many-body effect appears below the so-called pseudogap temperature $T^*$~\cite{Ohashi}.
In addition, the theoretical framework for the optical spin conductivity can be applied to other classes of topological superconductors and superfluidity such as 
spin-1/2 $s$-wave superfluid Fermi gas with the spin-orbit coupling~\cite{Sato2009,Toikka},
the $p$-wave superfluids in a Bose-Fermi mixture ~\cite{WuBruun,Zhu} and 
the spin Hall response in higher-dimensional systems with the chiral {\it p}-wave pairing symmetry.
It is worth investigating the spin conductance~\cite{Lebrat,Pace,Nakada,Ono,Sekino_mesospin,Ominato} detected by the quantum point contact in topological Fermi superfluids.

\acknowledgements
{The authors thank M. Matsuo, Y. Ominato and A. Furusaki for fruitful discussions.}
HT also thanks S. Tsutsui, T. M. Doi, and K. Iida for useful discussions in the initial stage of this study.
YS is supported by JSPS KAKENHI Grants No.\ 19J01006.
HT is supported by Grant-in-Aid for Scientific Research provided by JSPS through No.~18H05406.
SU is supported by MEXT Leading Initiative for Excellent Young Researchers,
JSPS KAKENHI Grant No.~JP21K03436,
and Matsuo Foundation.

\end{document}